\documentclass[journal]{IEEEtran}

\usepackage{url}
\usepackage{amsmath,amsthm,amssymb,amsbsy}
\usepackage{mathdots}
\usepackage{paralist}
\usepackage{xcolor}
\usepackage{color}
\usepackage{graphicx}
\usepackage{algorithm,algpseudocode}
\usepackage{comment}

\usepackage{fancyhdr}
\usepackage{cite}
\usepackage{cleveref}

\newtheorem{thm}{Theorem}
\newtheorem{cor}{Corollary}

\newtheorem{lem}{Lemma}
\theoremstyle{remark}

\def \Real {\mathbb{R}}
\newcommand{\C}{\mathbb{C}}

\newcommand{\N}{\mathbb{N}}

\newcommand{\vct}[1]{\boldsymbol{#1}}
\newcommand{\mtx}[1]{\boldsymbol{#1}}

\newcommand{\T}{\mathrm{T}}

\newcommand{\rank}{\operatorname{rank}}
\newcommand{\diag}{\operatorname{diag}}

\newcommand{\eps}{\epsilon}

\newcommand{\calS}{\mathcal{S}}
\newcommand{\calT}{\mathcal{T}}

\newcommand{\calB}{\mathcal{B}}

\newcommand{\va}{\vct{a}}

\newcommand{\ve}{\vct{e}}

\newcommand{\vx}{\vct{x}}
\newcommand{\vy}{\vct{y}}

\newcommand{\mB}{\mtx{B}}

\newcommand{\mD}{\mtx{D}}
\newcommand{\mE}{\mtx{E}}
\newcommand{\mF}{\mtx{F}}

\newcommand{\mL}{\mtx{L}}

\newcommand{\mU}{\mtx{U}}
\newcommand{\mV}{\mtx{V}}

\graphicspath{{./figs/}}

\newlength{\imgwidth}
\newboolean{twoColVersion}
\setboolean{twoColVersion}{false}
\newcommand{\twoCol}[2]{\ifthenelse{\boolean{twoColVersion}} {#1} {#2} }

\usepackage[english]{babel}
\title{The Eigenvalue Distribution of Discrete Periodic Time-Frequency Limiting Operators}
\author{Zhihui Zhu, Santhosh Karnik, Mark A.\ Davenport, Justin Romberg, Michael B. Wakin
\thanks{
This work was supported by NSF grants CCF-1409261 and  CCF-
1409406.}
\thanks{Z. Zhu and M. Wakin are with the Department
of Electrical Engineering, Colorado School of Mines, Golden, CO USA (e-mail: \{zzhu,mwakin\}@mines.edu).
S. Karnik, M. Davenport, and J. Romberg are with Electrical \& Computer Engineering, Georgia Institute of Technology, Atlanta, GA USA (e-mail:  \{skarnik1337,
mdav\}@gatech.edu, jrom@ece.gatech.edu).
}}

\begin{document}

\maketitle

\begin{abstract}
Bandlimiting and timelimiting operators play a fundamental role in analyzing bandlimited signals that are approximately timelimited (or vice versa). In this paper, we consider a time-frequency (in the discrete Fourier transform (DFT) domain) limiting operator whose eigenvectors are known as the {\em periodic discrete prolate
spheroidal sequences} (PDPSSs). We establish new nonasymptotic results on the eigenvalue distribution of this operator. As a byproduct, we also characterize the eigenvalue distribution of a set of submatrices of the DFT matrix, which is of independent interest.
\end{abstract}


\section{Introduction}
A series of seminal papers by Landau, Pollak, and Slepian explore the degree to which a bandlimited signal can be approximately timelimited~\cite{LandaP_ProlateII,LandaP_ProlateIII,SlepiP_ProlateI,Slepi_ProlateV}.
The key analysis involves a very special class
of functions---the {\em prolate spheroidal wave functions} (PSWF�s) in the continuous case and the {\em discrete prolate
spheroidal sequences} (DPSS�s) in the discrete case. These functions are the eigenvectors of the corresponding composition of bandlimiting and timelimiting operators and
provide a natural basis to use in a wide variety of applications involving bandlimiting and timelimiting~\cite{LandaP_ProlateII,LandaP_ProlateIII,SlepiP_ProlateI,Slepi_ProlateV,DavenSSBWB_Wideband,DavenportWakin2012CSDPSS,Zhu2015targetDetectDPSS}; see also~\cite{hogan2012duration} and the references therein for applications using PWSFs and see~\cite{ZhuWakin2015MDPSS} and the references therein for applications using DPSSs.

The {\em periodic discrete prolate
spheroidal sequences} (PDPSSs), introduced by Jain and Ranganath~\cite{jain1981extrapolation} and Gr{\"u}nbaum~\cite{grunbaum1981eigenvectors}, are the counterparts of the PSWFs in the finite dimensional case.  
The PDPSSs are the finite-length vectors whose discrete Fourier transform (DFT) is most concentrated in a given bandwidth.
Being simultaneously concentrated in the time and frequency domains makes these vectors useful in a number of signal processing applications. For example, Jain and Ranganath used PDPSSs for extrapolation and spectral estimation of periodic discrete-time signals~\cite{jain1981extrapolation} . PDPSSs were also used for limited-angle reconstruction in tomography~\cite{grunbaum1981eigenvectors }, for Fourier extension~\cite{matthysen2016fast}, and in~\cite{hogan2015wavelet}, the bandpass PDPSSs were used as a numerical approximation to the bandpass PSWFs for studying synchrony in sampled EEG signals.

The distribution of the eigenvalues of a time-frequency limiting operator dictate the (approximate) dimension of the space of signals which are bandlimited and approximately timelimited~\cite{SlepiP_ProlateI,Slepi_ProlateV}. Such distributions are known for the case of PSWFs and DPSSs. Specifically, an asymptotic expression for the PSWF eigenvalues was given in~\cite{SlepianCori}, and more recently, Israel~\cite{Israel2015EigenvalueDisTFLocalization} provided a non-asymptotic bound. Slepian~\cite{Slepi_ProlateV} first provided an asymptotic expression for the DPSS eigenvalues. In~\cite{Karnik2016FAST}, we recently provided a non-asymptotic result for the distribution of the DPSS eigenvalues (which improves upon a previous result in~\cite{ZhuWakin2015MDPSS}).

There exist comparatively few results concerning the PDPSS eigenvalues. In~\cite{xu1984periodic}, it was shown that unlike the PSWF and DPSS eigenvalues, the PDPSS eigenvalues can exactly achieve $0$ and $1$ and are degenerate in some cases. A non-asymptotic result on the distribution of the PDPSS eigenvalues was given in~\cite{EdelmanMT1998FutureFFT}. The special distribution of the PDPSS eigenvalues (See Figure~\ref{figure:eigPDPSS}) has been exploited for fast computing Fourier extensions of arbitrary extension length in~\cite{matthysen2016fast}.
In this paper, we provide a finer non-asymptotic result that improves upon the expression in~\cite{EdelmanMT1998FutureFFT}. We also characterize the spectrum of submatrices of the DFT matrix (see Corollary~\ref{cor:sub DFT block}), which is of independent interest in signal processing. For example, the low rank of DFT submatrices can be exploited for efficiently computing DFT~\cite{EdelmanMT1998FutureFFT}.

\section{Main Results}
\subsection{Time and Band-limiting Operators}
To begin, let $\calT_N:\C^M\rightarrow \C^M$ denote a time-limiting operator that only keeps the first $N\leq M$ entries of a vector, i.e., for any $\vx\in\C^M$,
\[
(\calT_N(\vx))[n] := \left\{\begin{array}{cc}\vx[n], & 0\leq n\leq N-1,\\
0, & N\leq n\leq M-1 . \end{array}\right.
\]
The DFT of any $\vx\in\C^M$, denoted by $\widehat \vx\in\C^M$, is defined as
\[
\widehat \vx[n]: = \frac{1}{\sqrt{M}}\sum_{m=0}^{M-1}\vx[m]e^{-j\frac{2\pi nm}{M}}, \ n\in[M],
\]
where $[M] = \left\{0,\ldots,M-1\right\}$.
Given $\widehat \vx$, $\vx$ can be recovered by taking the inverse DFT (IDFT), i.e.,
\[
\vx[m] = \frac{1}{\sqrt{M}}\sum_{n=0}^{M-1}\widehat\vx[n]e^{j\frac{2\pi nm}{M}}, \ m\in[M].
\]

Suppose $K\in\N$ such that $2K+1<M$. Let $\calB_K:\C^M\rightarrow \C^M$ denote a band-limiting operator that first zeros out the DFT of a vector outside the index range $\mathcal{I}_K := \{0,\ldots,K\}\cup\{M-K,\ldots,M-1\}$, then returns the corresponding signal in the time domain by taking the IDFT. That is
\begin{align*}
(\calB_K(\vx))[m] &:= \frac{1}{\sqrt{M}} \sum_{k \in \mathcal{I}_K} \widehat\vx[k]e^{j\frac{2\pi km}{M}}\\
&= \frac{1}{M}\sum_{k \in \mathcal{I}_K}\sum_{n=0}^{M-1}\vx[n]e^{-j\frac{2\pi nk}{M}}e^{j\frac{2\pi km}{M}}\\
&=\sum_{n=0}^{M-1}\frac{\sin\left((2K+1)(m-n)\pi/M\right)}{M\sin\left((m-n)\pi/M\right)}\vx[n].
\end{align*}
Denote $W=\frac{2K+1}{2M}<\frac{1}{2}$. Let $\mB_{M,W}\in\C^{M\times M}$ denote a prolate matrix with entries
\[
\mB_{M,W}[m,n] = \frac{\sin\left(2\pi W(m-n)\right)}{M\sin\left((m-n)\pi/M\right)}, \ m,n\in[M].
\]
Note that $\calB_K$ is equivalent to $\mB_{M,W}$, whose eigenvectors are given by the PDPSSs~\cite{jain1981extrapolation,grunbaum1981eigenvectors}.

Let $[\mB_{M,W}]_N\in\C^{N\times N}$ be the leading principal submatrix of $\mB_{M,W}$ constructed by removing the last $M-N$ rows and columns from $\mB_{M,W}$. Composing the time- and band-limiting operators, we obtain the linear operator $\calT_N\calB_K\calT_N:\C^M\rightarrow\C^M$, which has the same non-zero eigenvalues as $[\mB_{M,W}]_N$. Similar to the case for the DPSSs which can be obtained  efficiently and numerically stably by computing the eigenvectors of a tridiagonal matrix \cite{Slepi_ProlateV}, Gr{\"u}nbaum~\cite{grunbaum1981eigenvectors} showed that the prolate matrix $[\mB_{M,W}]_N$ commutes with a tridiagonal matrix,  providing a stable and reliable method for computing the PDPSSs.

\subsection{Eigenvalue Concentration}
In the rest of the paper, we assume $2K+1<M$, which is of practical interest for applications (e.g., ~\cite{jain1981extrapolation,grunbaum1981eigenvectors,hogan2015wavelet}).
Let $1\geq\lambda_N^{(0)}\geq \lambda_N^{(1)} \geq \cdots \lambda_N^{(N-1)}\geq 0$ denote the eigenvalues of $[\mB_{M,W}]_N$, where the upper and lower bounds follow because
\begin{align*}
\|\vx\|^2 \geq \vx^*\mB_{M,W}\vx = \sum_{k \in \mathcal{I}_K} |\widehat\vx[k]|^2 \geq 0
\end{align*}
for all $\vx \in \C^{M}$, indicating that the eigenvalues of $\mB_{M,W}$ are between 0 and 1 (and thus so are the eigenvalues of $[\mB_{M,W}]_N$ by the Sturmian separation theorem~\cite{Horn1985MatrixAnalysis}). We note that when $2K+1>M$, it is possible that some eigenvalues $\lambda_N^{(\ell)}>1$; see~\cite{xu1984periodic} for more discussion on this.

We establish the following results concerning the eigenvalue distribution for $[\mB_{M,W}]_N$.

\begin{figure}[htb!]
\centerline{
\includegraphics[width=4in]{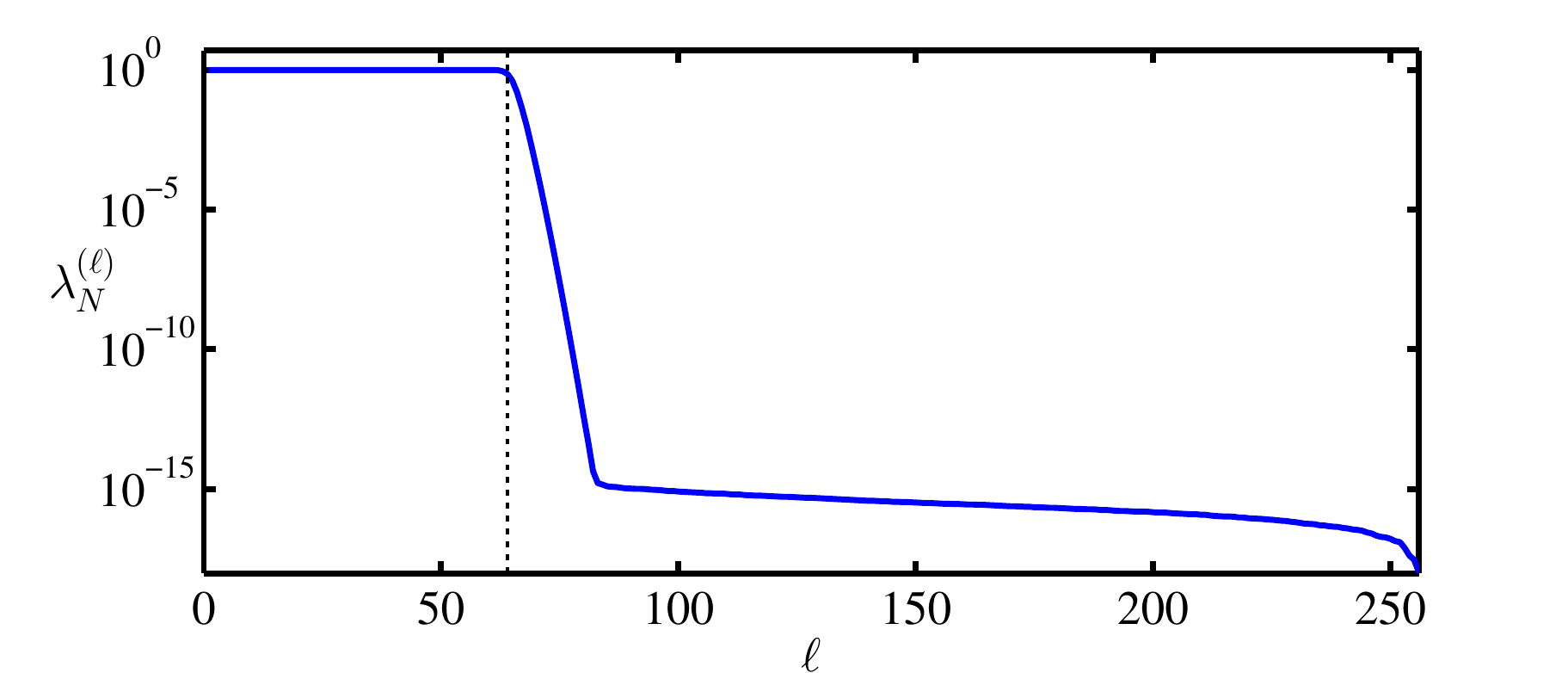}}
\caption{Eigenvalues of the prolate matrix $[\mB_{M,W}]_N$ with $M = 1024, N = 256, K = 128$ so that $\frac{(2K+1)N}{M}\approx 64$ (dashed line).}  \label{figure:eigPDPSS}
\end{figure}

\begin{figure}[htb!]
\centerline{
\includegraphics[width=4in]{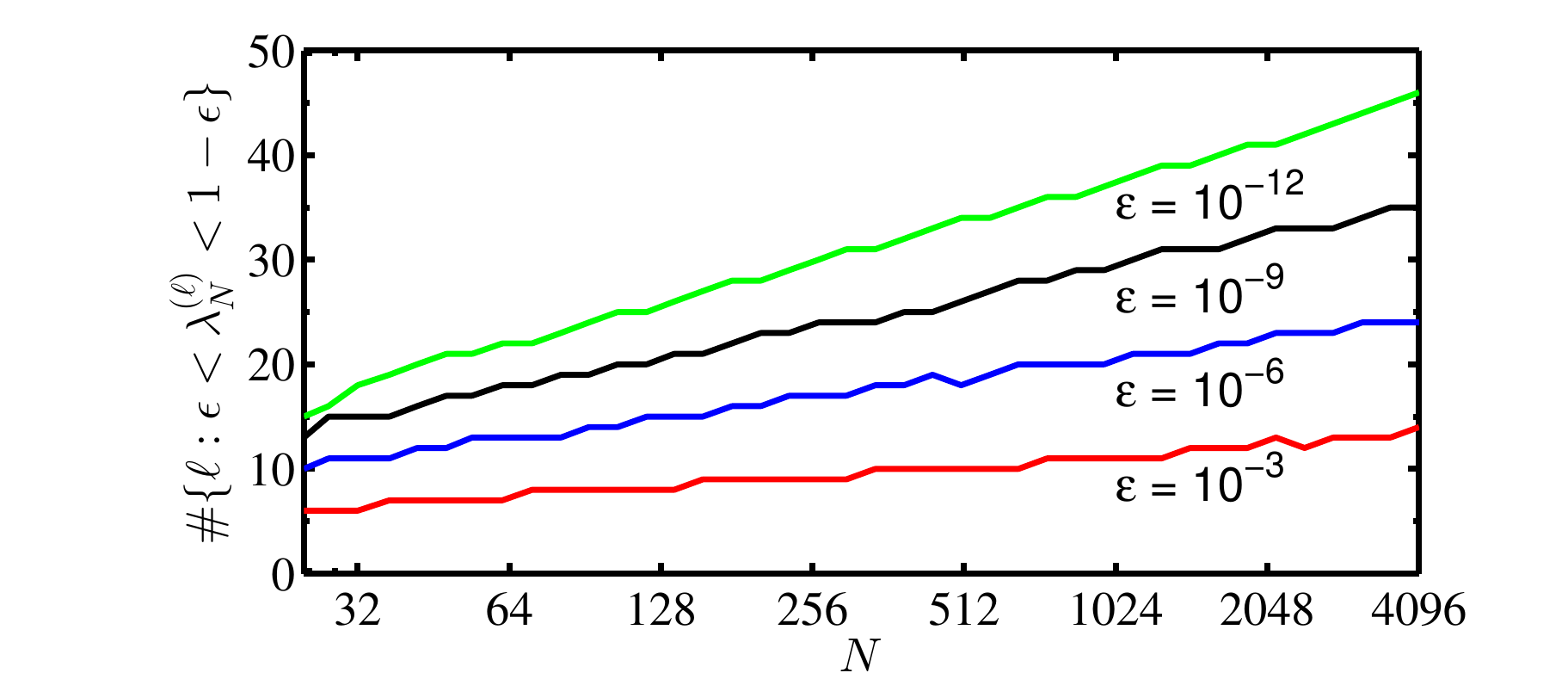}}
\caption{Width of the transition band $\#\{\ell:\epsilon<\lambda_N^{(\ell)}< 1-\epsilon\}$ for $N = \frac{1}{4}M, K = \frac{1}{8}M$, and $\epsilon = 10^{-3}, 10^{-6}, 10^{-9}, 10^{-12}$.}  \label{figure:Transition}
\end{figure}

\begin{thm}\label{thm:concentration of the PDPSS eigenvalues}(Spectrum concentration for $[\mB_{M,W}]_N$)
For any $M,N,K\in\N$, suppose $N<M$ and $W = \frac{2K+1}{2M}<\frac{1}{2}$.  Then for any $\epsilon\in(0,\frac{1}{2})$, we have
\[
\lambda_{N}^{(2\lfloor NW \rfloor- R(N,M,\epsilon))} \geq 1-\epsilon, \quad \lambda_{N}^{(2\lfloor NW \rfloor + R(N,M,\epsilon)+1)} \leq \epsilon,
\]
and
\[
\#\{\ell:\epsilon< \lambda_{N}^{(\ell)}< 1-\epsilon\} \leq 2R(N,M,\epsilon),
\]
where
\begin{align*}
R(N,M,\epsilon) &= \left(\frac{4}{\pi^2} \log(8N) + 6\right) \log \left( \frac{16}{\eps} \right) \\
& \quad \quad + 2\max\left(\frac{-\log \left( 8\pi\left(\left(\frac{M}{N}\right)^2-1\right) \epsilon \right)}{\log\left(\frac{M}{N}\right)},0\right). \end{align*}
\end{thm}
In words, Theorem \ref{thm:concentration of the PDPSS eigenvalues} implies that the first $\approx \frac{(2K+1)N}{M}$ eigenvalues tend to cluster very close to one, while the remaining eigenvalues tend to concentrate about zero, after a narrow transition band of width $O(\log\frac{1}{\epsilon}\log N)$.\footnote{$O(\cdot)$ denotes the standard ``big-O'' notation.} Figure~\ref{figure:eigPDPSS} presents an example to illustrate this phenomenon. We note that this phenomenon has been utilized in~\cite{matthysen2016fast} for efficiently computing Fourier extensions.
Figure~\ref{figure:Transition} shows the size of the eigenvalue gap $\#\{\ell:\epsilon< \lambda_{N}^{(\ell)}< 1-\epsilon\}$ for different $N$ and $\epsilon$, illustrating  that the size is proportional to $\log\frac{1}{\epsilon}\log N$.

Theorem \ref{thm:concentration of the PDPSS eigenvalues} also has implications regarding the distribution of singular values of submatrices of the DFT matrix.  Specifically, let $\mF_M$ be the normalized DFT matrix with entries given by
\[
\mF_M[m,n] = \frac{1}{\sqrt{M}}e^{-j2\pi\frac{mn}{M}},\ m,n\in[M].
\]
Let $L = \frac{M}{p}$ be an integer and let $\mF_{M|p}$ denote an $L\times L$ submatrix of $\mF_M$ obtained by deleting any consecutive $M-L$ columns and any consecutive $M-L$ rows of $\mF_M$.
Edelman et al.~\cite{EdelmanMT1998FutureFFT} proposed an approximate algorithm for DFT computations with lower communication cost based on the compressibility (low rank) of the blocks of $\mF_M$, i.e., $\mF_{M|p}$.  Let $1\geq\sigma^{(0)}\geq \sigma^{(1)}\geq \cdots\geq \sigma^{(L-1)}\geq 0$ denote the singular values of $\mF_{M|p}$. For any $\eps\in\left(0,\frac{1}{4}\right)$, a bound on the number of singular values such that $\sqrt\epsilon\leq \sigma^{(\ell)}\leq \sqrt{1-\epsilon}$ is given in~\cite{EdelmanMT1998FutureFFT}, which shows  $ \#\left\{\ell:\sqrt\epsilon\leq \sigma^{(\ell)}\leq \sqrt{1-\epsilon}\right\} \sim O(\log L)$. This bound highlights the logarithmic dependence on $L$, but ignores the dependence on\footnote{With simple manipulation, this bound states $O(1/\epsilon)$ dependence on $\epsilon$, which is quite large when $\epsilon$ is very small.} $\epsilon$. A finer non-asymptotic bound on the width of this transition region is given as follows.

\begin{cor}\label{cor:sub DFT block}
For any $M,p\in\N$ such that $L=\frac{M}{p}$ is an integer, let $\mF_{M|p}$ denote an $L \times L$ submatrix of the normalized DFT matrix $\mF_M$ obtained by deleting any consecutive $M-L$ columns and any consecutive $M-L$ rows of $\mF_M$. Let $\sigma^{(0)}\geq \sigma^{(1)}\geq \cdots\geq \sigma^{(L-1)}$ denote the singular values of $\mF_{M|p}$. Then for any $\eps\in(0,\frac{1}{2})$,
\[
\sigma^{(2\lfloor \frac{L}{2p}\rfloor - R(L,M,\eps))}\geq \sqrt{1-\eps}, \quad \sigma^{(2\lfloor \frac{L}{2p}\rfloor + R(L,M,\eps)+1)}\leq \sqrt{\eps},
\]
and
\[
\#\left\{\ell,\sqrt{\eps}<\sigma^{(\ell)}<\sqrt{1-\eps}\right\}\leq 2R\left(L,M,\eps\right),
\]
where $R(\cdot,\cdot,\cdot)$ is specified in Theorem~\ref{thm:concentration of the PDPSS eigenvalues}.
\end{cor}

\section{Proof of the main results}
\subsection{Supporting Results on DPSSs}
\label{sec:pre}
For any $N\in\N$ and $W\in(0,\frac{1}{2})$, define $\overline\mB_{N,W}\in\C^{N\times N}$ as a prolate matrix with elements
\[
\overline\mB_{N,W}[m,n] = \frac{\sin\left(2\pi W(m-n)\right)}{\pi (m-n)}
\]
for all $m,n\in\{0,\ldots,N-1\}$. Both $\overline\mB_{N,W}$ and $[\mB_{M,W}]_N$ are $N\times N$ Toeplitz matrices and they have very similar elements---the former has sinc elements while the latter has Dirichlet elements. Our idea for proving Theorem~\ref{thm:concentration of the PDPSS eigenvalues} is to utilize the spectrum concentration of $\overline\mB_{N,W}$ and exploit the similarity between $\overline\mB_{N,W}$ and $[\mB_{M,W}]_N$.

Let $\overline\lambda_{N,W}^{(\ell)}$ denote the eigenvalues of $\overline\mB_{N,W}$ placed in decreasing order. The DPSS vectors are the eigenvectors of $\overline\mB_{N,W}$~\cite{Slepi_ProlateV}. Let $\mF_{N,W}$ denote the partial normalized DFT matrix with the lowest $2\lfloor NW\rfloor+1$ frequency DFT vectors of length $N$, i.e.,
\[
\mF_{N,W} = \left[\begin{array}{ccc}\frac{1}{\sqrt N}\ve_{-\frac{\lfloor NW\rfloor}{N}}  & \cdots & \frac{1}{\sqrt N}\ve_{\frac{\lfloor NW\rfloor}{N}} \end{array}\right],
\]
where, for $f\in[-\frac{1}{2},\frac{1}{2}]$,
$$\ve_f:=\begin{bmatrix}e^{j2\pi f 0} & e^{j2\pi f 1} & \cdots & e^{j2\pi f (N-1)}\end{bmatrix}^\T \in \mathbb{C}^{N}$$
are the sampled exponentials.
The following result establishes that the difference between $\overline\mB_{N,W}$ and  $\mF_{N,W} \mF_{N,W}^*$ has an effective rank of $O(\log N\log\frac{1}{\eps})$.
\begin{lem}\cite[Theorem 1]{Karnik2016FAST}
Let $N \in \N$ and $W \in (0, \tfrac{1}{2})$ be given. Then for any $\eps \in (0,\tfrac{1}{2})$, there exist $N \times N$ matrices $\mL_1$ and $\mE_1$ such that
\[
\overline\mB_{N,W} = \mF_{N,W} \mF_{N,W}^* + \mL_1 + \mE_1,
\]
where
\[
\rank(\mL_1) \le \left( \frac{4}{\pi^2} \log(8N) + 6 \right) \log \left( \frac{15}{\eps} \right) ~~\text{and}~~ \|\mE_1\| \le \eps.
\]
\label{lem:factorization of B_NW}
\end{lem}

%

\subsection{Proof of Theorem~\ref{thm:concentration of the PDPSS eigenvalues}}
Lemma~\ref{lem:factorization of B_NW} implies that the difference between $\overline\mB_{N,W}$ and  $\mF_{N,W} \mF_{N,W}^*$ is effectively low rank. The main idea is to first show that the difference between the two prolate matrices $\overline\mB_{N,W}$ and $[\mB_{M,W}]_N$ is also effectively low rank.
By using the Taylor series
\[
\frac{1}{\sin x} - \frac{1}{x} = \sum_{r=1}^\infty \frac{2(1-2^{-(2r-1)})\zeta(2r)}{\pi^{2r}}x^{2r-1},
\]
where $\zeta$ denotes the Riemann-Zeta function, the $(m,n)$-th entry of the difference $[\mB_{M,W}]_N - \overline\mB_{N,W}$ is given by
\begin{align*}
&\left([\mB_{M,W}]_N - \overline\mB_{N,W}\right)[m,n] \\
&= \frac{\sin\left(2\pi W(m-n)\right)}{M\sin\left((m-n)\pi/M\right)} - \frac{\sin\left(2\pi W(m-n)\right)}{\pi (m-n)}\\
&= \left(\frac{1}{{\sin\left(\frac{(m-n)\pi}{M}\right)}} - \frac{1}{\frac{(m-n)\pi}{M}} \right)\frac{\sin\left(2\pi W(m-n) \right)}{M}\\
& = \sum_{r=1}^\infty t(r;m-n) = \mL_2[m,n] + \mE_2[m,n]
\end{align*}
for all $m,n=0,1,\ldots,N-1$. Here $
t(r;k) := \frac{2}{M\pi}\left(1-2^{1-2r} \right)\zeta(2r) \left(\frac{k}{M}\right)^{2r-1}\sin\left(2\pi W k \right)$, and
$\mL_2$ and $\mE_2$ are $N\times N$ matrices with entries
\begin{align*}
\mL_2[m,n]=  \sum_{r=1}^R t(r;m-n), \ \mE_2[m,n]= \sum_{r=R+1}^\infty t(r;m-n).
\end{align*}
Define $\mD\in\Real^{2R\times 2R}$ to have entries
\begin{align*}
\mD[2r-1-p,p] = \frac{2}{M\pi}\left(1 - 2^{-(2r-1)}\right)\zeta(2r)(-1)^p {2r-1 \choose p}
\end{align*}
for $1\leq r\leq R$ and $0\leq p\leq 2r-1$. Also define $\mU,\mV \in \Real^{N\times 2R}$ such that
\begin{align*}
\mU[n,r] = \left(\frac{n}{N}\right)^r\sin\left(2\pi Wn \right),\ \mV[n,r] = \left(\frac{n}{N}\right)^r\cos\left(2\pi Wn \right)
\end{align*}
for all $0\leq r \leq 2R-1$ and $0\leq n\leq N-1$. The rank of $\mL_2$ then can be identified by noting that
\begin{align*}
&\mL_2[m,n] =\sum_{r=1}^R\sum_{p=0}^{2r-1}\mD[2r-1-p,p]\left(\frac{m}{M}\right)^{2r-1-p}\left(\frac{n}{M}\right)^{p}
\cdot\\
&\quad \left(\sin\left(2\pi Wm\right)\cos\left(2\pi Wn\right) - \cos\left(2\pi Wm\right)\sin\left(2\pi Wn\right)    \right)\\
& = \sum_{r=1}^R\sum_{p=0}^{2r-1} \mD[2r-1-p,p]\big( \mU[m,2r-1-p]\mV[n,p] \\
&\qquad\qquad\qquad\qquad \qquad\qquad- \mV[m,2r-1-p]\mU[p,n]
\big)\\
&= \left(\mU\mD\mV^*\right)[m,n] - \left(\mV\mD\mU^*\right)[m,n].
\end{align*}
This implies $\mL_2 = \mU\mD\mV^* - \mV\mD\mU^*$ and $\rank(\mL_2)\leq 4R$.

Also note that $1-2^{1-s}\zeta(s) = \eta(s)$ is the Dirichlet eta function satisfying $
0<\eta(s) = \sum_{n=1}^\infty \frac{(-1)^{n-1}}{n^s}<1
$
for all $s\geq 1$. We now turn to bound the entries in $\mE_2$ as
\begin{align*}
\left|\mE_2(m,n)\right| &= \left| \sum_{r=R+1}^\infty t (r;m-n)\right|\leq \sum_{r=R+1}^\infty \frac{2}{\pi M}\left(\frac{N}{M}\right)^{2r-1}\\& = \frac{2}{\pi N} \frac{\left(\frac{N}{M}\right)^{2R+2}}{1-\left(\frac{N}{M}\right)^2}=\frac{2}{\pi N}\frac{\left(\frac{N}{M}\right)^{2R}}{\left(\frac{M}{N}\right)^2-1}.
\end{align*}
Choosing $R=\max \left( \frac{-\log 8\pi((\frac{M}{N})^2-1)\epsilon}{2\log \frac{M}{N}},0\right)$, we obtain that $\left|\mE_2(m,n)\right|\leq \frac{\epsilon}{16N}$. It follows from the Gershgorin circle theorem that
\begin{align*}
\left\|\mE_2\right\|\leq \max_{m}\sum_{n}\left|\mE_2(m,n)\right| \leq \frac{\eps}{16}.
\end{align*}

By Lemma~\ref{lem:factorization of B_NW}, there exist $N \times N$ matrices $\mL_1$  and $\mE_1$ with
 \[
\rank(\mL_1) \le \left( \tfrac{4}{\pi^2} \log(8N) + 6 \right) \log \left( \tfrac{16}{\eps} \right), \quad \|\mE_1\| \le \tfrac{15}{16}\eps
\]
such that $
\overline\mB_{N,W} = \mF_{N,W} \mF_{N,W}^* + \mL_1 + \mE_1.
$

Denoting $\mL = \mL_1 + \mL_2$ and $\mE = \mE_1 + \mE_2$, we obtain
\[
[\mB_{M,W}]_N = \overline\mB_{N,W} + \mL_2 + \mE_2 = \mF_{N,W} \mF_{N,W}^* + \mL + \mE,
\]
where
\begin{align*}
&\rank(\mL) \leq \rank(\mL_1) + \rank(\mL_2) \\
&\le \left( \tfrac{4}{\pi^2} \log(8N) + 6 \right) \! \log \left( \tfrac{16}{\eps} \right) + 2\max \!\left(\tfrac{-\log 8\pi((\frac{M}{N})^2-1)\epsilon}{\log(\frac{M}{N})},0\right) \end{align*}
and
\[ \|\mE\| \leq \|\mE_1\| + \|\mE_2\| \le \frac{15}{16}\eps + \frac{1}{16}\eps=\eps.
\]

Now we utilize the fact that $\mF_{N,W} \mF_{N,W}^*$ has only eigenvalues 1 and 0 to obtain the eigenvalue distribution for $[\mB_{M,W}]_N$. For all integers $\ell\in [N]$, the Weyl-Courant minimax representation of the eigenvalues can be stated as
\begin{align*}
\lambda_{N}^{(\ell)} &= \left\{\begin{array}{c}\min_{S_{\ell}}\max_{\|\vy\|_2=1,\vy\perp \calS_\ell} \left\langle [\mB_{M,W}]_N\vy,\vy\right\rangle,\\
\max_{\calS_{\ell+1}}\min_{\vy \in \calS_{\ell+1},\|\vy\|_2 =1}\left\langle [\mB_{M,W}]_N\vy,\vy\right\rangle,
\end{array}\right.
\end{align*}
where $\calS_{\ell}$ is an $\ell$-dimensional subspace of $\C^N$.

Recall that both $[\mB_{M,W}]_N$ and $\mL$ are symmetric, which implies that $\mE$ is also symmetric. Let $\widehat{\calS}$ be the  column space of $[\mF_{N,W} \quad \mL]$ and $\widehat d = \dim(\widehat{\calS})$.
\begin{align*}
\lambda_{N}^{(\widehat d)} = & \min_{\calS_{\widehat d}}\max_{\|\vy\|_2=1,\vy\perp \calS_{\widehat d}} \left\langle [\mB_{M,W}]_N\vy,\vy\right\rangle\\
\leq & \max_{\|\vy\|_2=1,\vy\perp \widehat{\calS}} \left\langle [\mB_{M,W}]_N\vy,\vy\right\rangle\\
=& \max_{\|\vy\|_2=1,\vy\perp \widehat{\calS}} \left\langle \left([\mB_{M,W}]_N -\mF_{N,W}\mF_{N,W}^*  -\mL\right)\vy,\vy\right\rangle\\
\leq & \|\mE\|\leq  \epsilon.
\end{align*}
We have $\widehat d = \dim(\widehat{\calS})\leq 2\lfloor NW \rfloor + R(N,M,\epsilon) + 1$, giving
\begin{align*}
\lambda_{N}^{(2\lfloor NW \rfloor + R(N,M,\epsilon)+1)} \leq \lambda_N^{(\widehat d)}\leq \eps.
\end{align*}
On the other hand, let $\overline \calS$ be the intersection of the column space of $\mF_{N,W}$ and the null space of $\mL$. Denote $\overline d = \dim(\overline{\calS})$.  We have
\begin{align*}
\lambda_{N}^{(\overline d-1)} = & \max_{\calS_{\overline d}}\min_{\vy \in \calS_{\overline d},\|\vy\|_2 =1}\left\langle [\mB_{M,W}]_N\vy,\vy\right\rangle\\
\geq & \min_{\vy \in \overline{\calS},\|\vy\|_2 =1}\left\langle [\mB_{M,W}]_N\vy,\vy\right\rangle\\
= &\min_{\vy \in \overline{\calS},\|\vy\|_2 =1}\left\langle \left([\mB_{M,W}]_N -\mF_{N,W}\mF_{N,W}^*  -\mL\right)\vy,\vy\right\rangle \\
& \quad\quad\quad\quad+ \left\langle \left(\mF_{N,W}\mF_{N,W}^*  +\mL\right)\vy,\vy\right\rangle\\
\geq & 1 -\|\mE\| \geq 1 - \epsilon.
\end{align*}
Noting that $\overline d\geq 2\lfloor NW \rfloor - R(N,M,\epsilon) + 1$, we obtain
\begin{align*}
\lambda_{N}^{\left(2\lfloor NW \rfloor - R(N,M,\epsilon)\right)} \geq \lambda_N^{(\overline d -1)}\geq 1-\eps.
\end{align*}

\subsection{Proof of Corollary~\ref{cor:sub DFT block}}
We first consider the spectrum of $[\mF_{M}]_L$, the top-left principal submatrix of $\mF_{M}$. It is clear that its singular values are between 0 and 1 since it is a submatrix of $\mF_{M}$. We first observe that the gram matrix of $[\mF_{M}]_L$, $[\mF_{M}]_L^*[\mF_{M}]_L$ is identical to $[\mB_{M,1/2p}]_L$ up to unitary phase factors, i.e.,
\begin{align*}
&\left([\mF_M]_L^*[\mF_M]\right)[m,n] = e^{-j\pi \frac{m-n}{L-1}}\frac{\sin\left(\pi (m-n)/p\right)}{M\sin\left((m-n)\pi/M\right)}\\
&= e^{-j\pi \frac{m-n}{L-1}}\left([\mB_{M,1/2p}]_L\right)[m,n], \ \forall \ m,n\in[L].
\end{align*}
This implies $[\mF_M]_L^*[\mF_M]_L$ has the same eigenvalue distribution to $[\mB_{M,1/2p}]_L$. Thus, Corollary~\ref{cor:sub DFT block} holds for $[\mF_M]_L$ trivially by following Theorem \ref{thm:concentration of the PDPSS eigenvalues}.

Now note that any submatrix $\mF_{M|p}$ obtained by deleting any consecutive $M-L$ columns and any consecutive $M-L$ rows of $\mF_M$ is identical to $[\mF_M]_L$ up to unitary phase factors
\begin{align*}
\mF_{M|p}= \diag\left(\va_\xi\right) [\mF_M]_L \diag\left(\va_\eta\right),
\end{align*}
where $\xi,\eta$ depend on the locations of the submatrix $\mF_{M|p}$ and
\[
\va_{\xi} = \begin{bmatrix}1 &  e^{-2\pi \frac{\xi}{M}} & \cdots & e^{-2\pi \frac{(L-1)\xi}{M}} \end{bmatrix}^\T.
\]
Thus, any submatrix $\mF_{M|p}$ has the same spectrum as $[\mF_M]_L$.

\bibliographystyle{ieeetr}
\bibliography{bibfileFAST}

\end{document}